\documentstyle[aps,fleqn]{revtex}
\newcounter{subeq}

\begin{document}
\draft
\title{Connected Green function approach to ground state symmetry
breaking in $\Phi^4_{1+1}$-theory\footnote{supported by DFG, BMFT, KFA
J\"ulich and GSI Darmstadt}\footnote{part of the dissertation of
A. Peter}}
\author{J.M. H\"auser, W. Cassing, A. Peter and M.H. Thoma\\
Institut f\"ur Theoretische Physik, Universit\"at Gie\ss en \\
35392 Gie\ss en, Germany \\}
\maketitle
\date{\today}
\begin{abstract}
Using the cluster expansions for n-point Green functions
we derive a closed set of dynamical equations of motion for connected
equal-time Green functions by neglecting all connected functions
higher than $4^{th}$ order for the $\lambda \Phi^4$-theory in $1+1$
dimensions. We apply the equations to the investigation of spontaneous
ground state symmetry breaking, i.e. to the evaluation of the
effective potential at temperature $T=0$.
Within our momentum space discretization
we obtain a second order phase transition (in agreement with the
Simon-Griffith theorem) and a critical coupling of
$\lambda_{crit}/4m^2=2.446$ as compared to a first order phase
transition and
$\lambda_{crit}/4m^2=2.568$ from the Gaussian effective potential
approach.
\end{abstract}
\section{Introduction}
\label{introduction}
In the last years there have been many attempts to develop
non-perturbative
methods for quantum field theoretical (QFT)
problems such as the infrared behaviour of QCD.
Many of the non-perturbative methods used so far,
such as the coupled cluster expansion \cite{1,2,3},
originate from standard many-body theory and use states in the Fock
representation, thus they have the general problem of Fock
representations
being disjunct to the Hilbert space of QFT \cite{3,4}.

In this work we also adopt a successful method from standard many-body
theory, which however is formulated in terms of Green functions and
therefore of a genuine field-theoretical nature. We propose an
approach
along the line of n-body correlation dynamics \cite{5},
which describes the propagation of the system in terms of equal-time
Green functions (i.e. we use the density matrix formalism).
The basic strategy of correlation dynamics is the truncation of the
infinite hierarchy of equations of motion for n-point Green functions
via
the use of cluster expansions, i.e. the expansion of Green functions
in
terms of connected Green functions (correlation functions).
In this work we go up to the connected 4-point level, i.e. we include
the 2-,3- and 4-field correlation functions.
For relativistic systems with a nonlocal interaction our method
does not guarantee a covariant description, since retardation effects
can
not be included due to the equal-time formalism.
There is, however, no objection against the use of an
equal-time formalism in the case of a local relativistic field theory
as
long as one consistently considers Green functions containing
canonically conjugate field momenta as well as the fields themselves.

This paper is organized as follows: In section \ref{correldyn} we
describe
the derivation of the correlation dynamical equations of motion
for $\Phi^4$-theory in $1+1$ dimensions; the
final set of equations itself is shifted to appendix \ref{eqoms} in
view
of its length. Section \ref{application} is devoted to the application
of the method to the evaluation of the effective potential at zero
temperature within various limits and thus to the investigation of
ground
state symmetry breaking.
\section{Correlation dynamics for $\Phi^4_{1+1}$-theory}
\label{correldyn}
We consider the Lagrangian density
\begin{eqnarray}
{\cal L} = \frac{1}{2} \partial_\mu \phi \partial^\mu \phi
- \frac{1}{2} m_0^2 \phi^2 - \frac{1}{4} \lambda \phi^4 \; ,
\label{lagrangian}
\end{eqnarray}
in one space and one time dimension, which corresponds to the
Hamiltonian
\begin{eqnarray}
H=\frac{1}{2} \int dx \left[ \pi^2
+ \left( \frac{\partial \phi}{\partial x} \right)^2
+ m_0^2 \phi^2 + \frac{1}{2} \lambda \phi^4 \right] \; ,
\label{hamiltonian}
\end{eqnarray}
where $\pi=\partial_t \phi$ and $m_0$ is the bare mass of the scalar
field
$\phi$.
In order to evaluate the effective potential of the theory at zero
temperature, i.e. the minimum of the energy density for a given
magnetization
$\langle \phi \rangle$, we decompose $\phi$ into a classical and a
quantum
part according to
\begin{eqnarray}
\phi=\Phi_0 + \Phi \; ,
\label{splitoff}
\end{eqnarray}
where $\Phi_0$ is a real constant and assume
\begin{eqnarray}
\langle \Phi \rangle = 0 \; , \; \; {\rm i.e.} \; \;
\langle \phi \rangle = \Phi_0 \; .
\label{splitoffassumption}
\end{eqnarray}
Since $\Phi_0$ is a constant, we have
\begin{eqnarray}
\pi=\partial_t \phi=\partial_t \Phi = \Pi \; .
\label{piassumption}
\end{eqnarray}
The Hamiltonian, expressed in terms of the classical part $\Phi_0$
and the
quantum part $\Phi$ of $\phi$, then reads
\begin{eqnarray}
\lefteqn{
H = \frac{1}{2} \int dx \left[ \Pi^2
+ \left( \frac{\partial \Phi}{\partial x} \right) ^2
+ \left( 2 \Phi_0 m_0^2 + 2 \lambda \Phi_0^3 \right) \Phi
+ \left( m_0^2 + 3 \lambda \Phi_0^2 \right) \Phi^2
+ 2 \lambda \Phi_0 \Phi^3 \right. } \nonumber \\
& & \left. + \frac{1}{2} \lambda \Phi^4
+ m_0^2 \Phi_0^2 + \frac{1}{2} \lambda \Phi_0^4 \right] \; .
\label{splitoffhamiltonian}
\end{eqnarray}
For the time evolution of $\Phi$ and $\Pi$ we obtain with
(\ref{splitoffhamiltonian}) and the canonical equal-time commutation
relations
\begin{eqnarray}
\left[ \Phi(x,t),\Phi(y,t) \right]
= \left[ \Pi(x,t),\Pi(y,t) \right] =0 \; , \; \;
\left[ \Phi(x,t), \Pi(y,t) \right] = i \delta (x-y)
\label{commurel}
\end{eqnarray}
by means of the Heisenberg equation:
\begin{eqnarray*}
\partial_t \Phi = \Pi \; ,
\end{eqnarray*}
\begin{eqnarray}
\partial_t \Pi = - \left( \Phi_0 m_0^2 + \lambda \Phi_0^3 \right)
+ \left( \partial_x^2 - m_0^2 - 3\lambda \Phi_0^2 \right) \Phi
- 3 \lambda \Phi_0 \Phi^2
- \lambda \Phi^3 \; .
\label{heisenbergequations}
\end{eqnarray}
In analogy to (\ref{heisenbergequations}) we get the equations of
motion
for equal-time operator products of $\Phi$ and $\Pi$, e.g.
\begin{eqnarray*}
\partial_t \left( \Phi(x_1,t) \Phi(x_2,t) \right)  = \Pi (x_1,t)
\Phi (x_2,t)
+ \Phi (x_1,t) \Pi (x_2,t) \; ,
\end{eqnarray*}
\begin{eqnarray}
\lefteqn{
\partial_t \left( \Pi(x_1,t) \Phi(x_2,t) \right) =
\left[ - \left( \Phi_0 m_0^2 + \lambda \Phi_0^3 \right)
+ \left( \partial_x^2 - m_0^2 - 3\lambda \Phi_0^2 \right) \Phi(x_1,t)
\right.
} \nonumber \\
& & \left. - 3 \lambda \Phi_0 \Phi^2(x_1,t)
- \lambda \Phi^3(x_1,t) \right] \Phi(x_2,t)
+ \Pi(x_1,t) \Pi(x_2,t) \; , \; \; ... \; \; .
\label{hierarchy}
\end{eqnarray}
After taking their expectation values, the equations of motion for all
possible n-point equal-time products of $\Phi$ and $\Pi$
comprise an infinite hierarchy of equations of motion for equal-time
Green
functions. The analogon of this in nonrelativistic many-body theory
is the BBGKY\footnote{{\bf B}orn {\bf B}ogoliubov {\bf G}reen
{\bf K}irkwood {\bf Y}von} density matrix hierarchy \cite{6}.
Since we are considering a local field theory, the equations of motion
for
the Green functions contain no retardation integrals, which
implies that working in an equal-time limit is sufficient for
describing the
propagation of the system as long as one considers all Green functions
containing the fields as well as their conjugate momenta.

For practical purposes, the infinite hierarchy of coupled differential
equations of first order in time has to be truncated. This is done
using
the cluster expansions for n-point Green functions \cite{7},
i.e. their decomposition
into sums of products of connected Green functions, which works as
long as
the system is in a pure phase \cite{8,9}.
The explicit form of the cluster expansions can be derived
from the generating functionals of full and connected Green functions,
$Z[J, \sigma]$ and $W[J, \sigma]$, given by
\begin{eqnarray}
Z[J, \sigma] = {\rm Tr} \left\{ \rho \; T \left[ e^{i \int d^2
\hat{x}
\left(
J(\hat{x}) \Phi(\hat{x}) + \sigma(\hat{x}) \Pi(\hat{x}) \right) }
\right] \right\}
\; \; \; \; {\rm and} \; \; \; \;
Z[J, \sigma] = e^{ W[J, \sigma] } \; ,
\end{eqnarray}
respectively,
where $T$ is the time ordering operator, $\rho$ is the statistical
density
operator describing the pure or mixed state of the system
(${\rm Tr} \rho=1$) and $\hat{x}=(x,t)$.
We start with the cluster expansions for the time-ordered Green
functions
with different time arguments:
\begin{eqnarray}
{\langle}\Phi(\hat{x}) {\rangle} \;=\; {\langle}\Phi(\hat{x})
{\rangle}_c
\; \; , \; \; \; \;
{\langle}\Pi(\hat{x}) {\rangle} \;=\; {\langle}\Pi(\hat{x})
{\rangle}_c \; ,
\end{eqnarray}
\begin{eqnarray}
\lefteqn{
{\langle}T \Phi(\hat{x}_1) \Phi(\hat{x}_2){\rangle} =
\lim_{J, \sigma \to 0}
\frac{\delta}{i\delta J(\hat{x}_1)}
\frac{\delta}{i\delta J(\hat{x}_2)} e^{W[J, \sigma]}
} \nonumber \\
& & = \lim_{J, \sigma \to 0}
\frac{\delta}{i\delta J(\hat{x}_1)} \left\{ \left(
\frac{\delta}{i\delta J(\hat{x}_2)} W[J, \sigma] \right)
e^{W[J, \sigma]}
\right\}
\nonumber \\
& & = \lim_{J, \sigma \to 0} \left\{ \left(
\frac{\delta}{i\delta J(\hat{x}_1)}
\frac{\delta}{i\delta J(\hat{x}_2)} W[J, \sigma] \right)
+ \left( \frac{\delta}{i\delta J(\hat{x}_1)} W[J, \sigma] \right)
\left( \frac{\delta}{i\delta J(\hat{x}_2)} W[J, \sigma] \right)
\right\}
e^{W[J, \sigma]}
\nonumber \\
\nonumber \\
& & = {\langle}T \Phi(\hat{x}_1) \Phi(\hat{x}_2){\rangle}_c +
{\langle}\Phi(\hat{x}_1){\rangle}{\langle}\Phi(\hat{x}_2)
{\rangle} \; ,
\end{eqnarray}
where $\langle \cdot \rangle_c$ denotes the connected part of the
expectation value.
Analogously we obtain
\begin{eqnarray*}
\langle T \Pi(\hat{x}_1) \Phi(\hat{x}_2) \rangle = \langle
T \Pi(\hat{x}_1)
\Phi(\hat{x}_2) \rangle_c
+ \langle \Pi(\hat{x}_1) \rangle \langle \Phi(\hat{x}_2) \rangle \; ,
\end{eqnarray*}
\begin{eqnarray*}
\langle T \Pi(\hat{x}_1) \Pi(\hat{x}_2) \rangle = \langle
T \Pi(\hat{x}_1)
\Pi(\hat{x}_2) \rangle_c
+ \langle \Pi(\hat{x}_1) \rangle \langle \Pi(\hat{x}_2) \rangle \; ,
\end{eqnarray*}
\begin{eqnarray*}
\lefteqn{
{\langle}T \Phi(\hat{x}_1) \Phi(\hat{x}_2) \Phi(\hat{x}_3)
{\rangle} \; = \;
{\langle}T \Phi(\hat{x}_1) \Phi(\hat{x}_2) \Phi(\hat{x}_3)
{\rangle}_c
+ {\langle}T \Phi(\hat{x}_1) \Phi(\hat{x}_2){\rangle}_c
{\langle}\Phi(\hat{x}_3)
{\rangle}
} \nonumber \\
& & + {\langle}T \Phi(\hat{x}_1) \Phi(\hat{x}_3){\rangle}_c
{\langle}\Phi(\hat{x}_2){\rangle}
+ {\langle}T \Phi(\hat{x}_2) \Phi(\hat{x}_3) {\rangle}_c
{\langle}\Phi(\hat{x}_1){\rangle}
+ {\langle}\Phi(\hat{x}_1){\rangle} {\langle}\Phi(\hat{x}_2){\rangle}
{\langle}\Phi(\hat{x}_3){\rangle}
\; , \; \; ... \; \; .
\end{eqnarray*}
\begin{eqnarray}
\end{eqnarray}
The expressions for equal-time Green functions are obtained by
taking the
well-defined equal-time limit which yields the appropriate
operator ordering
in the cluster expansions. We arrive at
\begin{eqnarray*}
{\langle}\Phi(x) {\rangle} \;=\; {\langle}\Phi(x){\rangle}_c
\; \; , \; \; \; \;
{\langle}\Pi(x) {\rangle} \;=\; {\langle}\Pi(x){\rangle}_c \; ,
\end{eqnarray*}
\begin{eqnarray*}
{\langle} \Phi(x_1) \Phi(x_2){\rangle}
= {\langle} \Phi(x_1) \Phi(x_2){\rangle}_c + {\langle}\Phi(x_1)
{\rangle}{\langle}\Phi(x_2){\rangle} \; ,
\end{eqnarray*}
\begin{eqnarray*}
\langle  \Pi(x_1) \Phi(x_2) \rangle = \langle  \Pi(x_1) \Phi(x_2)
\rangle_c
+ \langle \Pi(x_1) \rangle \langle \Phi(x_2) \rangle \; ,
\end{eqnarray*}
\begin{eqnarray*}
\langle  \Pi(x_1) \Pi(x_2) \rangle = \langle  \Pi(x_1) \Pi(x_2)
\rangle_c
+ \langle \Pi(x_1) \rangle \langle \Pi(x_2) \rangle \; ,
\end{eqnarray*}
\begin{eqnarray*}
\lefteqn{
{\langle} \Phi(x_1) \Phi(x_2) \Phi(x_3) {\rangle} \; = \;
{\langle} \Phi(x_1) \Phi(x_2) \Phi(x_3) {\rangle}_c
+ {\langle} \Phi(x_1) \Phi(x_2){\rangle}_c {\langle}\Phi(x_3){\rangle}
} \nonumber \\
& & + {\langle} \Phi(x_1) \Phi(x_3){\rangle}_c {\langle}\Phi(x_2)
{\rangle}
+ {\langle} \Phi(x_2) \Phi(x_3) {\rangle}_c {\langle}\Phi(x_1)
{\rangle}
+ {\langle}\Phi(x_1){\rangle} {\langle}\Phi(x_2){\rangle}
{\langle}\Phi(x_3){\rangle}
\; , \; \; ... \; \; ,
\end{eqnarray*}
\begin{eqnarray}
\end{eqnarray}
where all (equal) time arguments have been suppressed.
In view of their length the cluster expansions for the other Green
functions required for our calculations are not explicitly given here,
but e.g. can be found in \cite{10}.

Due to equations (\ref{splitoffassumption}) and (\ref{piassumption})
the 1-point functions in all cluster expansions are now assumed to
vanish,
which in the end leads to the same result as
considering the cluster expansions for Green functions containing the
original fields $\phi$ and $\pi$ and setting
$\langle \phi \rangle=\Phi_0$
and $\langle \pi \rangle=0$.

The cluster expansions then are truncated by neglecting all
connected n-point
Green functions with ${\rm n} > {\rm N}$; in our case with
${\rm n} > 4$.
Inserting the truncated cluster expansions into the equations
of motion
of type (\ref{hierarchy}) up to the equations for the 4-point
functions leads
to a closed system of coupled nonlinear equations for the
connected Green
functions (i.e. a system of correlation dynamical equations)
up to the 4-point level in analogy to the n-body correlation
dynamics in nonrelativistic many-body theory \cite{5}.
For this straightforward but somewhat tedious derivation,
which in view of
its length is not explicitly given here,
all truncated cluster expansions up to the 6-point
level are required, since the highest order Green functions
appearing in
the hierarchy equations (\ref{hierarchy}) up to the 4-point level
are the 6-point functions.

The resulting equations of motion for the connected Green
functions still
have to be renormalized. $\Phi^4$-theory in $1+1$ dimensions is
superrenormalizable and only requires a mass renormalization. There is
only one mass counterterm due to the divergent tadpole diagram
comprising
the contribution to the selfenergy of lowest order in the coupling
constant.
This mass counterterm can be evaluated analytically either in the
framework
of perturbation theory or, equivalently, by normal ordering the
Hamiltonian with respect to the perturbative
vacuum\footnote{In addition to introducing the mass counterterm, one
of course has to subtract the constant infinite zero-point
contribution
from the Hamiltonian, which however does not enter the Heisenberg
equations.}. We obtain \cite{11}:
\begin{eqnarray}
m_0^2=m^2+\delta m^2
\; \; , \; \; \; \;
\delta m^2= -3 \lambda \Delta (0)
\label{counterterm}
\end{eqnarray}
with
\begin{eqnarray}
\Delta(x_1-x_2)=\int \frac{dp}{2 \pi}
\frac{e^{ip(x_1-x_2)}}{2 \omega(p)}
\; \; , \; \; \; \;
\omega(p)=\sqrt{p^2+m^2} \; .
\label{capdeltadefinition}
\end{eqnarray}
The logarithmically divergent mass counterterm (\ref{counterterm})
can be
analytically removed from the equations of motion by normal
ordering all
operator products with respect to the perturbative vacuum and thereby
splitting off the short-distance singularities of the free equal-time
Green functions. The normal ordering within the connected Green
functions
only takes place within the 2-point functions, since in the other
connected
n-point functions all field operators commute with each other.
For the 2-point functions we have:
\begin{eqnarray*}
\langle \Phi(x_1) \Phi(x_2) \rangle_c
= \langle : \Phi(x_1) \Phi(x_2) : \rangle_c
+ \Delta(x_1 - x_2) \; ,
\end{eqnarray*}
\begin{eqnarray*}
\langle \Pi(x_1) \Pi(x_2) \rangle_c
= \langle : \Pi(x_1) \Pi(x_2) : \rangle_c
- \left( \partial_{x_1}^2 - m^2 \right) \Delta(x_1-x_2) \; ,
\end{eqnarray*}
\begin{eqnarray*}
\langle \Pi(x_1) \Phi(x_2) \rangle_c
= \langle : \Pi(x_1) \Phi(x_2) : \rangle_c
- \frac{i}{2} \delta(x_1-x_2) \; ,
\end{eqnarray*}
\begin{eqnarray}
\langle \Phi(x_1) \Pi(x_2) \rangle_c
= \langle : \Phi(x_1) \Pi(x_2) : \rangle_c
+ \frac{i}{2} \delta(x_1-x_2) \; .
\label{normalorder}
\end{eqnarray}
After inserting (\ref{counterterm}) and (\ref{normalorder}) into the
equations of motion for the connected Green functions, all terms
containing a factor $\Delta(0)$ mutually cancel out, leading to a
closed set of renormalized equations for the normal ordered connected
Green functions.

The final step is to transform the renormalized correlation dynamical
equations of motion from coordinate space to an arbitrary single
particle
basis in order to simplify their numerical integration.
In this respect we expand the field $\Phi$ and its conjugate
momentum $\Pi$ according to
\begin{eqnarray}
\Phi(x)=\sum_{\alpha} \Phi_\alpha \psi_\alpha(x)
\; \; , \; \; \; \;
\Pi(x)=\sum_{\alpha} \Pi_\alpha \psi_\alpha(x)
\; \; \; \;
{\rm with}
\; \; \; \;
\int dx \; \psi^*_\alpha(x) \psi_\beta(x)=\delta_{\alpha \beta} \; .
\label{operatorexpansion}
\end{eqnarray}
For the corresponding equal-time Green functions we then have
\begin{eqnarray*}
\lefteqn{
\langle \Phi(x_1) \Phi(x_2) \rangle_c
= \sum_{\alpha \beta} \langle \Phi_\alpha \Phi_\beta \rangle_c
\psi_\alpha(x_1) \psi_\beta(x_2)
= \langle :\Phi(x_1) \Phi(x_2): \rangle_c +\Delta(x_1-x_2)
} \nonumber \\
& & = \sum_{\alpha \beta} \langle :\Phi_\alpha \Phi_\beta: \rangle_c
\psi_\alpha(x_1) \psi_\beta(x_2)
+ \sum_{\alpha \beta} \Delta_{\alpha \beta}
\psi_\alpha(x_1) \psi_\beta(x_2) \; ,
\end{eqnarray*}
\begin{eqnarray*}
\langle \Phi(x_1) \Phi(x_2) \Phi(x_3) \rangle
= \sum_{\alpha \beta \gamma}
\langle \Phi_\alpha \Phi_\beta \Phi_\gamma \rangle
\psi_\alpha(x_1) \psi_\beta(x_2) \psi_\gamma(x_3) \; ,
\end{eqnarray*}
\begin{eqnarray}
\langle \Phi(x_1) \Phi(x_2) \Phi(x_3) \rangle_c
= \sum_{\alpha \beta \gamma}
\langle \Phi_\alpha \Phi_\beta \Phi_\gamma \rangle_c
\psi_\alpha(x_1) \psi_\beta(x_2) \psi_\gamma(x_3)
\; \; , \; \; \; \;
... \; \; .
\label{greenfunctionexpansions}
\end{eqnarray}
The corresponding expressions for the other Green functions are
obtained
analogously.
By inserting the expansions (\ref{greenfunctionexpansions}) into the
renormalized equations of motion and projecting out the matrix
elements with
respect to the single particle basis we obtain the final result for
the
correlation dynamical equations of motion for the
$\Phi^4_{1+1}$-theory,
which are
denoted by $\Phi^4CD$ ($\Phi^4$ $C$orrelation $D$ynamics) furtheron,
and can directly be used for a numerical integration.
In view of their length the $\Phi^4CD$
equations are shifted to appendix \ref{eqoms}.
\section{Application to ground state symmetry breaking}
\label{application}
In this section we apply the $\Phi^4CD$ equations to the
determination of
the effective potential of the $\Phi^4_{1+1}$-theory at zero
temperature\footnote{i.e. the ground state energy density
in a subspace
with a given magnetization} and
thereby investigate the spontaneous breakdown of the symmetry under
the discrete transformation $\phi \to -\phi$ for values of the
coupling
exceeding a critical value, which manifests itself in a
nonzero ground state magnetization $\langle \phi \rangle$.

Since we also want to study the influence of the different connected
n-point functions, we introduce 4 limiting cases of the
correlation dynamical equations; these are denoted as $\Phi^4CD(2)$,
$\Phi^4CD(2,3)$, $\Phi^4CD(2,4)$ and $\Phi^4CD(2,3,4)$, where the
numbers
in parantheses are the orders of connected n-point functions that are
taken into account (i.e., for instance in the $\Phi^4CD(2,4)$ case the
connected 3-point functions are set equal to zero and
$\Phi^4CD(2,3,4)$
denotes the original $\Phi^4CD$ approximation discussed in section
\ref{correldyn}).

In order to integrate equations (\ref{firsteqom}) - (\ref{lasteqom})
numerically, we choose plane waves in a one-dimensional box with
periodic
boundary conditions as a single particle basis, i.e. we work in
discretized momentum space. In order to select an appropriate box
size
for a given renormalized mass, we compare the numerically obtained
GEP\footnote{{\bf G}aussian {\bf E}ffective {\bf P}otential} solution
for the ground state configuration within the discretized
system to the
analytically accessible GEP solution in the continuum limit (for
the GEP approximation, see appendix \ref{gep} or
\cite{11,12,13}). Due to the large amount of computer time needed
for $\Phi^4CD$ calculations we have to
compromise between a good momentum space resolution and the
convergence
with a minimum number of plane waves.
It has proven to be most effective to choose a box size of 100 fm for
a renormalized mass of 10 MeV. In fig. \ref{picphstran0} we show the
ground state magnetization $\langle \phi \rangle=\Phi_0$
as a function of the dimensionless coupling $\lambda/4m^2$ for these
parameters, where different numbers of single particle states have
been
taken into account. The vertical line shows the position of the
critical
coupling for symmetry breaking in the continuum limit. For practical
purposes we will always use the 15 lowest lying plane waves in the
following
calculations, unless explicitly stated otherwise. Within the GEP
approximation we then obtain a critical coupling of
$\lambda_{crit}/4m^2=2.568$ for the discretized system  as
compared to a
critical coupling of $\lambda_{crit}/4m^2=2.5527045$ in the
continuum limit.

Since the $\Phi^4CD$ equations only describe the propagation
of equal-time
Green functions for a given initial configuration, they cannot
directly be
applied to the evaluation of static equilibrium properties of
the system.
As in the case of correlation dynamics for nonrelativistic
many-body theory
there is no easy access to the stationary solutions of the
equations, and
moreover it is not clear, in how
far additional constraints have to be
imposed on the subspace of stationary configurations in order to
select only
the physical solutions.

We therefore use a different approach for the
evaluation of equal-time
ground state Green functions within the $\Phi^4CD$ approximation.
Starting with the trivial exact ground state configuration for
$\lambda=0$
and a given fixed value of $\Phi_0$ as an initial condition,
we continuously
switch on the coupling while propagating the system in time.
The time-dependence of the coupling is chosen to be linear with
$\lambda=\beta t$ and $\beta=const$.

In fig. \ref{picconv} the energy density obtained within this
time-dependent
method is shown as a function of the coupling for different values of
$\beta$ for $\Phi_0=0.4$ and all 4 limiting cases of the correlation
dynamical equations. In all cases an asymptotic curve is approached
with
decreasing $\beta$, i.e. when the coupling is switched on more
slowly.
This indicates that in the limit
$\beta \to 0$
the whole process becomes fully adiabatic, i.e. the system will
time-dependently follow the trajectory of the ground state as a
function
of the coupling. For the $\Phi^4CD(2)$ approximation,
taking into account only the connected 2-point functions, which
is the field
theoretical analogon to time-dependent Hartree-Fock theory, we
have direct
access to the static ground state solutions
since its stationary limit
is simply given by the GEP approximation.
The corresponding curves are
shown in the upper left part of fig. \ref{picconv}; the GEP is
displayed as the lower one of the two solid lines. Indeed, the
asymptotic
curve of the time-dependent method is identical to the GEP solution.
In addition, we have checked numerically that for all 4 limiting
cases
of correlation dynamics the system increasingly equilibrates along
its trajectory with decreasing $\beta$, i.e. if we stop increasing
$\lambda$
at some point in time, the system will remain in its present
state when propagated further.

In choosing a finite value for $\beta$ we have to compromise between
a good convergence of our time-dependent method and a minimum of
computer
time we want to invest; the curves in the two following pictures have
therefore been evaluated with $\beta=5 {\rm MeV}^2 {\rm c}/{\rm fm}$.

The ground state energy density for various values of $\Phi_0$
is plotted versus the coupling $\lambda/4m^2$
in fig. \ref{picinccpl} for all 4 limiting
cases of correlation dynamics,
where for the reasons mentioned above the $\Phi^4CD(2)$ approximation
has
been replaced by the GEP approximation. The GEP and the
$\Phi^4CD(2,4)$
approximation each predict a first order phase transition,
since the first
curve to intersect the $\Phi_0=0$ (i.e. the symmetric phase)
energy density
in the GEP case has $\Phi_0=0.6$ and in the
$\Phi^4CD(2,4)$ case has
$\Phi_0=0.8$; i.e. they both lead to a finite value of $\Phi_0$,
implying
that the vacuum magnetization has to jump to that value
discontinuously
at the critical value of the coupling constant (compare fig.
\ref{picphstran0} for the GEP).
In contrast to that, the $\Phi^4CD(2,3)$ approximation and the
$\Phi^4CD(2,3,4)$ approximation each predict a second order phase
transition,
since in both cases all curves with $\Phi_0 \ne 0$ intersect the
curve with $\Phi_0=0$ in the correct order,
thus enabling the vacuum magnetization to increase
continuously once the critical coupling is exceeded.

At this point it is useful to recall, that there is a rigorous
mathematical
proof of the statement that there can be no first order phase
transition
in the $\Phi^4_{1+1}$-theory (in the continuum limit) \cite{14}. This
proof is based on the Simon-Griffith theorem \cite{15},
which in turn is
obtained by considering the $\Phi^4$ field theory as a proper limit of
a generalized Ising model.

Thus we conclude, that in order to describe the order of the phase
transition
correctly, the inclusion of the connected 3-point function is
required.
A simple ''geometrical'' explanation of this fact will be given
in the discussion below.

In fig. \ref{picveff} the effective potential is plotted as a function
of the magnetization $\Phi_0$ for various values of the coupling. The
same data have been used as for the previous figure. In this
representation,
the first order nature of the phase transition in the GEP and the
$\Phi^4CD(2,4)$ case can be seen from the fact, that for these
approximations
there is a maximum in the effective potential between the minimum at
$\Phi_0=0$ and the second minimum on the right hand side
(for the GEP this
minimum is not very pronounced); thus at the critical coupling
the second minimum has to be located at a finite value of $\Phi_0$.
In the case of the $\Phi^4CD(2,3)$ and the $\Phi^4CD(2,3,4)$
approximation
there is no such intermediate maximum, and as the critical coupling is
approached from above, the second minimum is shifted to the
left towards $\Phi_0=0$.

For $\Phi_0 \to \infty$ the system will approach the classical limit,
i.e. with decreasing $\Phi_0$ higher order correlations increasingly
become
important; the inclusion of correlations always lowers the energy
density, since a larger configuration space is opened up for the
system.
However, for obvious mathematical reasons the inclusion of the 2-point
function alone cannot lower the energy density at $\Phi_0=0$; unlike
the
other n-point functions of even order the 2-point function therefore
is most
important in the region on the right hand side in the plots of fig.
\ref{picveff}. The 4-point function, as the highest order correlation
included in this work, is most important around $\Phi_0=0$, i.e. for
small classical field strengths. The intermediate maximum now
appears in
between the domains governed by the 2-point function on the
right and the
4-point function on the left, an inclusion of higher
correlations of even
order would most probably lower the energy density in a
region even more
concentrated around $\Phi_0=0$ and therefore not cure this problem.
The inclusion of connected n-point functions of uneven order cures the
problem, since for symmetry reasons these have to vanish as
$\Phi_0 \to 0$.
The 3-point function therefore obviously assumes its main importance
exactly in the region in between the domains of the 2- and the 4-point
function, it ''bends down'' the intermediate maximum.

Finally, in table \ref{table} we give the values for the critical
coupling
extracted from our calculations for different numerical parameters
and the
4 different limiting cases of $\Phi^4CD$.
The critical couplings $\lambda_{crit}/4m^2$ obtained by other
authors,
that are also using non-perturbative techniques adopted from
standard many-body theory, are as follows:
\cite{16,17} obtain 1.829 or 1.375, respectively, using the
discretized
light-front quantization method,
\cite{18} obtain 1.72 using a varitional approximation with
trial states
which are quartic rather than gaussians (quadratic exponentials)
causing an inclusion of the 3-field correlation amplitude in the
language of the coupled cluster method \cite{1},
\cite{2} obtain the estimate $0.95<\lambda/4m^2<2.15$
using an improved version of the coupled cluster expansion method,
and using second order perturbation theory in the residual interaction
leads to 1.14 \cite{19}.
All of these authors obtain a second order phase transition in
agreement
with the Simon-Griffith-theorem.
\begin{table}[b]
\begin{tabular}{|c|c|c|c|c|}
\hline
approximation & No. of states \quad \quad \quad \quad & $\beta$
[${\rm MeV}^2 {\rm c}/{\rm fm}$] \quad \quad \quad \quad &
$\lambda_{crit}/4m^2$ \quad \quad \quad & order of phasetransition
\quad \quad \quad \\ \hline\hline
GEP & 15 & - & 2.568 & first \\ \hline
& 11 & 5.0 & 1.679 & \\ \cline{2-4}
& 13 & 5.0 & 1.649 & \\ \cline{2-4}
$\Phi^4CD$(2,3) & 15 & 5.0 & 1.629 & second \\ \cline{2-4}
& 17 & 5.0 & 1.613 & \\ \cline{2-4}
& 19 & 5.0 & 1.601 & \\ \hline
$\Phi^4CD$(2,4) & 15 & 5.0 & 3.81 & first \\ \hline
$\Phi^4CD$(2,3,4) & 15 & 5.0 & 2.446 & second \\ \hline
\end{tabular}
\vspace{1cm}
\caption{\label{table} Values for the critical coupling for different
approximations and numerical parameters.}
\end{table}
\section{Summary and outlook}
\label{summary}
This work comprises the first application of correlation dynamics for
equal-time Green functions to a field theoretical problem, i.e. to the
determination of the effective potential in $\Phi^4_{1+1}$-theory.
After
giving a derivation of the corresponding equations of motion, we
showed
that we are able to evaluate equal-time quantities in the interacting
ground state for a given vacuum magnetization by time-dependently
increasing
the coupling in an adiabatic process.

Our numerical results predict a second order phase transition in
agreement
with the Simon-Griffith theorem, as soon as the connected 3-point
function
is included. However, since the connected 2-point function and the
connected
3-point function alone are not able to lower the energy of the
symmetric
phase, the critical coupling obtained within the $\Phi^4CD(2,3)$
approximation is too low. Going one step further in our expansion, we
find that the connected 4-point function in the $\Phi^4CD(2,3,4)$
method
increases the coupling to a value, which is only slightly lower than
the value obtained by the GEP approximation, where however the shape
of the effective potential changes completely as compared to the
GEP.

In general, the results of this work demonstrate the applicability of
correlation dynamics to the description of low-energy (ground state)
phenomena in local field theories.
Since in principle the equations are designed to describe the
propagation of the system in time with arbitrary initial conditions,
our method is also a potentially powerful tool for the investigation
of
non-equilibrium properties of relativistic field theories, e.g. the
response to external perturbations.

The present study can be viewed as a first step towards
the application of correlation dynamics to SU(N) gauge theories,
aiming at a non-perturbative description of the infrared behaviour of
QCD; we already presented the corresponding equations of motion in
\cite{10}.
\begin{appendix}
\section{Equations of motion of $\Phi^4$ correlation dynamics}
\label{eqoms}
In this appendix we present the renormalized correlation dynamical
equations
for the normal ordered connected Green functions up to the 4-point
level,
formulated with respect to an arbitrary single particle basis set
(see section \ref{correldyn}). In order to compactify the equations we
introduce the following abbreviations:
\begin{eqnarray*}
\langle \alpha | \lambda_1 \lambda_2 \rangle
= \int dx \; \psi^*_\alpha (x) \psi_{\lambda_1}(x)
\psi_{\lambda_2}(x) \; ,
\end{eqnarray*}
\begin{eqnarray*}
\langle \alpha | \lambda_1 \lambda_2 \lambda_3 \rangle
= \int dx \; \psi^*_\alpha (x) \psi_{\lambda_1}(x)
\psi_{\lambda_2}(x) \psi_{\lambda_3}(x) \; ,
\end{eqnarray*}
\begin{eqnarray*}
\langle \alpha \beta \gamma | 1 \rangle
= \int dx \; \psi^*_\alpha(x) \psi^*_\beta(x) \psi^*_\gamma(x)
\; ,
\end{eqnarray*}
\begin{eqnarray*}
\langle \alpha \beta \gamma | \lambda \rangle
= \int dx \; \psi^*_\alpha(x) \psi^*_\beta(x) \psi^*_\gamma(x)
\psi_\lambda(x) \; ,
\end{eqnarray*}
\begin{eqnarray*}
t_{\alpha \beta}=\int dx \; \psi^*_\alpha (x)
\left( \partial_x^2-m^2-3\lambda\Phi_0^2 \right) \psi_\beta(x) \; ,
\end{eqnarray*}
\begin{eqnarray*}
U_{\alpha \beta}=-3\lambda \sum_{\lambda_1 \lambda_2}
\langle \alpha | \lambda_1 \lambda_2 \beta \rangle
\langle : \Phi_{\lambda_1} \Phi_{\lambda_2} : \rangle_c \; .
\end{eqnarray*}
The permutation operator interchanging the indices $\alpha$ and
$\beta$
is denoted by ${\cal P}_{\alpha \beta}$. Although normal ordering only
affects the connected 2-point functions, we write out the normal
ordering
operation $: \cdot :$ in all connected Green functions
in order to have
a uniform notation. The equations then read:
%
%
\begin{eqnarray}
\frac{d}{dt} \langle: \Phi_\alpha \Phi_\beta  :\rangle_c =
\langle: \Pi_\alpha \Phi_\beta  :\rangle_c + \langle: \Phi_\alpha
\Pi_\beta
:\rangle_c
\; ,
\label{firsteqom}
\end{eqnarray}
\\
\begin{eqnarray}
\lefteqn{\frac{d}{dt} \langle: \Pi_\alpha \Phi_\beta  :\rangle_c =
\langle: \Pi_\alpha \Pi_\beta  :\rangle_c}
\nonumber\\ &&
+ \sum_{\lambda} ( t_{\alpha\lambda} + U_{\alpha\lambda})
\langle: \Phi_\lambda \Phi_\beta :\rangle_c
+ \sum_{\lambda} ( U_{\alpha\lambda}
- 3 \lambda \Phi_0^2 \delta_{\alpha\lambda}) \Delta_{\lambda\beta}
\nonumber\\ &&
- \lambda \sum_{\lambda_1\lambda_2\lambda_3}  \langle \alpha
|\lambda_1\lambda_2\lambda_3 \rangle
\langle: \Phi_{\lambda_1} \Phi_{\lambda_2} \Phi_{\lambda_3}
\Phi_{\beta} :\rangle_c
\nonumber\\ &&
- 3 \lambda \Phi_0 \sum_{\lambda_1\lambda_2} \langle \alpha
|\lambda_1 \lambda_2 \rangle
\langle: \Phi_{\lambda_1} \Phi_{\lambda_2} \Phi_{\beta} :\rangle_c
\; ,
\end{eqnarray}
\\
\begin{eqnarray}
\lefteqn{\frac{d}{dt} \langle: \Pi_\alpha \Pi_\beta :\rangle_c =
(1+{\cal P}_{\alpha\beta}) \sum_{\lambda}
(t_{\alpha\lambda} + U_{\alpha\lambda})
\langle: \Phi_{\lambda} \Pi_{\beta} :\rangle_c}
\nonumber\\ &&
- \lambda (1+{\cal P}_{\alpha\beta})
\sum_{\lambda_1\lambda_2\lambda_3}
\langle \alpha |\lambda_1\lambda_2\lambda_3 \rangle
\langle: \Phi_{\lambda_1} \Phi_{\lambda_2}
\Phi_{\lambda_3} \Pi_{\beta} :\rangle_c
\nonumber\\ &&
- 3 \lambda  \Phi_0 (1+{\cal P}_{\alpha\beta})
\sum_{\lambda_1 \lambda_2}
\langle \alpha |\lambda_1 \lambda_2 \rangle
\langle: \Phi_{\lambda_1} \Phi_{\lambda_2} \Pi_{\beta} :\rangle_c
\; ,
\end{eqnarray}
\\
%
%
\begin{eqnarray}
\frac{d}{dt} \langle: \Phi_\alpha \Phi_\beta \Phi_\gamma :\rangle_c =
  \langle: \Pi_\alpha \Phi_\beta \Phi_\gamma :\rangle_c
+ \langle: \Phi_\alpha \Pi_\beta \Phi_\gamma :\rangle_c
+ \langle: \Phi_\alpha \Phi_\beta \Pi_\gamma :\rangle_c
\; ,
\end{eqnarray}
\\
\begin{eqnarray}
\lefteqn{\frac{d}{dt} \langle: \Pi_\alpha \Phi_\beta \Phi_\gamma
:\rangle_c =
  \langle: \Pi_\alpha \Pi_\beta \Phi_\gamma :\rangle_c
+ \langle: \Pi_\alpha \Phi_\beta \Pi_\gamma :\rangle_c }
\nonumber\\ &&
+\sum_{\lambda} (t_{\alpha\lambda} + U_{\alpha\lambda})
\langle: \Phi_\lambda \Phi_\beta \Phi_\gamma :\rangle_c
\nonumber\\ &&
- 3 \lambda \sum_{\lambda_1 \lambda_2 \lambda_3}
\langle \alpha|\lambda_1\lambda_2\lambda_3\rangle
(1+{\cal P}_{\beta\gamma})
\langle:\Phi_{\lambda_1}\Phi_{\lambda_2} \Phi_{\beta}:\rangle_c
(\langle:\Phi_{\lambda_3}\Phi_{\gamma}:\rangle_c +
\Delta_{\lambda_3\gamma})
\nonumber\\ &&
-3\lambda \Phi_0 \sum_{\lambda_1\lambda_2}
\langle\alpha|\lambda_1\lambda_2\rangle \lbrace
\langle:\Phi_{\lambda_1}\Phi_{\lambda_2}\Phi_\beta
\Phi_\gamma:\rangle_c
\nonumber\\ &&
+ 2(\langle:\Phi_{\lambda_1}\Phi_\beta:\rangle_c +
\Delta_{\lambda_1\beta})
(\langle:\Phi_{\lambda_2}\Phi_\gamma:\rangle_c +
\Delta_{\lambda_2\gamma})
\rbrace
\; ,
\end{eqnarray}
\\
\begin{eqnarray}
\lefteqn{\frac{d}{dt} \langle: \Pi_\alpha \Pi_\beta
\Phi_\gamma :\rangle_c =
  \langle: \Pi_\alpha \Pi_\beta \Pi_\gamma :\rangle_c }
\nonumber\\ &&
+ (1+{\cal P}_{\alpha\beta})
\sum_{\lambda} (t_{\alpha\lambda} + U_{\alpha\lambda})
\langle: \Phi_\lambda \Pi_\beta \Phi_\gamma :\rangle_c
\nonumber\\ &&
- 3 \lambda (1+{\cal P}_{\alpha\beta}) \sum_{\lambda_1 \lambda_2
\lambda_3}
\langle \alpha|\lambda_1\lambda_2\lambda_3\rangle \lbrace
\langle:\Phi_{\lambda_1}\Phi_{\lambda_2} \Pi_{\beta}:\rangle_c
(\langle:\Phi_{\lambda_3}\Phi_{\gamma}:\rangle_c +
\Delta_{\lambda_3\gamma})
\nonumber\\ &&
+\langle:\Phi_{\lambda_1} \Phi_{\lambda_2} \Phi_\gamma:\rangle_c
 \langle:\Phi_{\lambda_3} \Pi_{\beta}:\rangle_c
\rbrace
\nonumber\\ \nonumber\\ &&
-3\lambda \Phi_0 (1+{\cal P}_{\alpha\beta})\sum_{\lambda_1\lambda_2}
\langle\alpha|\lambda_1\lambda_2\rangle \lbrace
\langle:\Phi_{\lambda_1}\Phi_{\lambda_2}\Pi_\beta\Phi_\gamma:\rangle_c
\nonumber\\ &&
+ 2 \langle:\Phi_{\lambda_1}\Pi_\beta:\rangle_c
(\langle:\Phi_{\lambda_2}\Phi_\gamma:\rangle_c +
\Delta_{\lambda_2\gamma})
\rbrace
\; , \end{eqnarray}
\\
\begin{eqnarray}
\lefteqn{\frac{d}{dt} \langle: \Pi_\alpha \Pi_\beta \Pi_\gamma
:\rangle_c =
(1+{\cal P}_{\alpha\beta}+{\cal P}_{\alpha\gamma})
\sum_{\lambda} (t_{\alpha\lambda} + U_{\alpha\lambda})
\langle: \Phi_\lambda \Pi_\beta \Pi_\gamma :\rangle_c }
\nonumber\\ &&
- 3 \lambda (1+{\cal P}_{\alpha\beta}+{\cal P}_{\alpha\gamma})
\sum_{\lambda_1 \lambda_2 \lambda_3}
\langle \alpha|\lambda_1\lambda_2\lambda_3\rangle
(1+{\cal P}_{\beta\gamma})
\langle:\Phi_{\lambda_1}\Phi_{\lambda_2} \Pi_{\beta}:\rangle_c
\langle:\Phi_{\lambda_3}\Pi_{\gamma}:\rangle_c
\nonumber\\ &&
-3\lambda \Phi_0 (1+{\cal P}_{\alpha\beta}+{\cal P}_{\alpha\gamma})
\sum_{\lambda_1\lambda_2}
\langle\alpha|\lambda_1\lambda_2\rangle \lbrace
\langle:\Phi_{\lambda_1}\Phi_{\lambda_2}\Pi_\beta\Pi_\gamma:\rangle_c
+ 2 \langle:\Phi_{\lambda_1}\Pi_\beta:\rangle_c
    \langle:\Phi_{\lambda_2}\Pi_\gamma:\rangle_c
\rbrace
\nonumber\\ &&
+ \frac{3}{2} \lambda\Phi_0
 \langle\alpha\beta\gamma|1\rangle
\; , \end{eqnarray}
\\
%
%
\begin{eqnarray}
\lefteqn{\frac{d}{dt}\langle:\Phi_\alpha \Phi_\beta \Phi_\gamma
\Phi_\delta :\rangle_c =
 \langle:\Pi_\alpha \Phi_\beta \Phi_\gamma \Phi_\delta :\rangle_c}
 \nonumber\\ &&
+\langle:\Phi_\alpha \Pi_\beta \Phi_\gamma \Phi_\delta :\rangle_c
+\langle:\Phi_\alpha \Phi_\beta \Pi_\gamma \Phi_\delta :\rangle_c
+\langle:\Phi_\alpha \Phi_\beta \Phi_\gamma \Pi_\delta :\rangle_c
\; , \end{eqnarray}
\\
\begin{eqnarray}
\lefteqn{\frac{d}{dt}\langle:\Pi_\alpha \Phi_\beta \Phi_\gamma
\Phi_\delta :\rangle_c =
 \langle:\Pi_\alpha \Pi_\beta \Phi_\gamma \Phi_\delta :\rangle_c
+\langle:\Pi_\alpha \Phi_\beta \Pi_\gamma \Phi_\delta :\rangle_c
+\langle:\Pi_\alpha \Phi_\beta \Phi_\gamma \Pi_\delta :\rangle_c}
\nonumber\\ &&
+\sum_{\lambda} (t_{\alpha\lambda} + U_{\alpha\lambda})
\langle:\Phi_\lambda\Phi_\beta\Phi_\gamma\Phi_\delta:\rangle_c
\nonumber\\ &&
-3\lambda \sum_{\lambda_1\lambda_2\lambda_3}
\langle\alpha|\lambda_1\lambda_2\lambda_3\rangle \lbrace
2(\langle:\Phi_{\lambda_1}\Phi_\beta:\rangle_c +
\Delta_{\lambda_1\beta})
(\langle:\Phi_{\lambda_2}\Phi_\gamma:\rangle_c +
\Delta_{\lambda_2\gamma})
(\langle:\Phi_{\lambda_3}\Phi_\delta:\rangle_c +
\Delta_{\lambda_3\delta})
\nonumber\\ &&
+(1+{\cal P}_{\beta\gamma}+{\cal P}_{\beta\delta})
[(\langle:\Phi_{\lambda_1}\Phi_\beta:\rangle_c +
\Delta_{\lambda_1\beta})
\langle:\Phi_{\lambda_2}\Phi_{\lambda_3}
\Phi_\gamma\Phi_\delta:\rangle_c
\nonumber\\ &&
+\langle:\Phi_{\lambda_1}\Phi_{\lambda_2}\Phi_\beta:\rangle_c
 \langle:\Phi_{\lambda_3}\Phi_\gamma\Phi_{\delta}:\rangle_c ]
\rbrace
\nonumber\\ \nonumber\\ &&
-6\lambda\Phi_0 \sum_{\lambda_1\lambda_2}
\langle\alpha|\lambda_1\lambda_2\rangle
(1+{\cal P}_{\beta\gamma}+{\cal P}_{\beta\delta})
(\langle:\Phi_{\lambda_1}\Phi_\beta:\rangle_c +
\Delta_{\lambda_1\beta})
\langle:\Phi_{\lambda_2}\Phi_\gamma\Phi_\delta:\rangle_c
\; , \end{eqnarray}
\\
\begin{eqnarray}
\lefteqn{\frac{d}{dt}\langle:\Pi_\alpha \Pi_\beta \Phi_\gamma
\Phi_\delta :\rangle_c =
 \langle:\Pi_\alpha \Pi_\beta \Pi_\gamma \Phi_\delta :\rangle_c
+\langle:\Pi_\alpha \Pi_\beta \Phi_\gamma \Pi_\delta :\rangle_c}
\nonumber\\ &&
+(1+{\cal P}_{\alpha\beta})
\sum_{\lambda} (t_{\alpha\lambda} + U_{\alpha\lambda})
\langle:\Phi_\lambda\Pi_\beta\Phi_\gamma\Phi_\delta:\rangle_c
\nonumber\\ &&
-3\lambda (1+{\cal P}_{\alpha\beta})
\sum_{\lambda_1\lambda_2\lambda_3}
\langle\alpha|\lambda_1\lambda_2\lambda_3\rangle \lbrace
2 \langle:\Phi_{\lambda_1}\Pi_\beta:\rangle_c
(\langle:\Phi_{\lambda_2}\Phi_\gamma:\rangle_c +
\Delta_{\lambda_2\gamma})
(\langle:\Phi_{\lambda_3}\Phi_\delta:\rangle_c +
\Delta_{\lambda_3\delta})
\nonumber\\ &&
+\langle:\Phi_{\lambda_1}\Pi_\beta:\rangle_c
\langle:\Phi_{\lambda_2}\Phi_{\lambda_3}\Phi_\gamma
\Phi_\delta:\rangle_c
+(1+{\cal P}_{\gamma\delta})
(\langle:\Phi_{\lambda_1}\Phi_{\gamma}:\rangle_c +
\Delta_{\lambda_1\gamma})
\langle:\Phi_{\lambda_2}\Phi_{\lambda_3}\Pi_\beta\Phi_\delta:\rangle_c
\nonumber\\ &&
+(1+{\cal P}_{\beta\gamma}+{\cal P}_{\beta\delta})
\langle:\Phi_{\lambda_1}\Phi_{\lambda_2}\Pi_\beta:\rangle_c
 \langle:\Phi_{\lambda_3}\Phi_\gamma\Phi_{\delta}:\rangle_c
\rbrace
\nonumber\\ \nonumber\\ &&
-6\lambda\Phi_0 (1+{\cal P}_{\alpha\beta}) \sum_{\lambda_1\lambda_2}
\langle\alpha|\lambda_1\lambda_2\rangle \lbrace
\langle:\Phi_{\lambda_1}\Pi_\beta:\rangle_c
\langle:\Phi_{\lambda_2}\Phi_\gamma\Phi_\delta:\rangle_c
\nonumber\\ &&
+(1+{\cal P}_{\gamma\delta})
(\langle:\Phi_{\lambda_1}\Phi_\gamma:\rangle_c +
\Delta_{\lambda_1\gamma})
\langle:\Phi_{\lambda_2}\Pi_{\beta}\Phi_{\delta}:\rangle_c
\rbrace
\; , \end{eqnarray}
\\
\begin{eqnarray}
\lefteqn{\frac{d}{dt}\langle:\Pi_\alpha \Pi_\beta \Pi_\gamma
\Phi_\delta :\rangle_c =
 \langle:\Pi_\alpha \Pi_\beta \Pi_\gamma \Pi_\delta :\rangle_c }
\nonumber\\ &&
+(1+{\cal P}_{\alpha\beta}+{\cal P}_{\alpha\gamma})
\sum_{\lambda} (t_{\alpha\lambda} + U_{\alpha\lambda})
\langle:\Phi_\lambda\Pi_\beta\Pi_\gamma\Phi_\delta:\rangle_c
\nonumber\\ &&
-3\lambda (1+{\cal P}_{\alpha\beta}+{\cal P}_{\alpha\gamma})
\sum_{\lambda_1\lambda_2\lambda_3}
\langle\alpha|\lambda_1\lambda_2\lambda_3\rangle \lbrace
2 \langle:\Phi_{\lambda_1}\Pi_\beta:\rangle_c
\langle:\Phi_{\lambda_2}\Pi_\gamma:\rangle_c
(\langle:\Phi_{\lambda_3}\Phi_\delta:\rangle_c +
\Delta_{\lambda_3\delta})
\nonumber\\ &&
+(1+{\cal P}_{\beta\gamma})
\langle:\Phi_{\lambda_1}\Phi_{\lambda_2}\Pi_\beta
\Phi_\delta:\rangle_c
\langle:\Phi_{\lambda_3}\Pi_\gamma:\rangle_c
+\langle:\Phi_{\lambda_1}\Phi_{\lambda_2}\Pi_\beta\Pi_\gamma:\rangle_c
(\langle:\Phi_{\lambda_3}\Phi_{\delta}:\rangle_c +
\Delta_{\lambda_3\delta})
\nonumber\\ \nonumber\\ &&
+(1+{\cal P}_{\beta\gamma}+{\cal P}_{\beta\delta})
\langle:\Phi_{\lambda_1}\Phi_{\lambda_2}\Pi_\beta:\rangle_c
 \langle:\Phi_{\lambda_3}\Pi_\gamma\Phi_{\delta}:\rangle_c
\rbrace
\nonumber\\ \nonumber\\ &&
-6\lambda\Phi_0 (1+{\cal P}_{\alpha\beta}+{\cal P}_{\alpha\gamma})
\sum_{\lambda_1\lambda_2}
\langle\alpha|\lambda_1\lambda_2\rangle \lbrace
(1+{\cal P}_{\beta\gamma})
\langle:\Phi_{\lambda_1}\Pi_\beta:\rangle_c
\langle:\Phi_{\lambda_2}\Pi_\gamma\Phi_\delta:\rangle_c
\nonumber\\ &&
+\langle:\Phi_{\lambda_1}\Pi_{\beta}\Pi_{\gamma}:\rangle_c
(\langle:\Phi_{\lambda_2}\Phi_\delta:\rangle_c +
\Delta_{\lambda_2\delta})
\rbrace
\nonumber\\ &&
+\frac{3}{2}\lambda \sum_{\lambda}
\langle\alpha\beta\gamma|\lambda\rangle
(\langle:\Phi_\lambda\Phi_\delta:\rangle_c + \Delta_{\lambda\delta})
\; , \end{eqnarray}
\\
\begin{eqnarray}
\lefteqn{\frac{d}{dt}\langle:\Pi_\alpha \Pi_\beta \Pi_\gamma
\Pi_\delta
:\rangle_c =
(1+{\cal P}_{\alpha\beta}+{\cal P}_{\alpha\gamma}+
{\cal P}_{\alpha\delta})
\sum_{\lambda} (t_{\alpha\lambda} + U_{\alpha\lambda})
\langle:\Phi_\lambda\Pi_\beta\Pi_\gamma\Pi_\delta:\rangle_c}
\nonumber\\ &&
-3\lambda
(1+{\cal P}_{\alpha\beta}+{\cal P}_{\alpha\gamma}+
{\cal P}_{\alpha\delta})
\sum_{\lambda_1\lambda_2\lambda_3}
\langle\alpha|\lambda_1\lambda_2\lambda_3\rangle \lbrace
2\langle:\Phi_{\lambda_1}\Pi_\beta:\rangle_c
 \langle:\Phi_{\lambda_2}\Pi_\gamma:\rangle_c
 \langle:\Phi_{\lambda_3}\Pi_\delta:\rangle_c
\nonumber\\ &&
+(1+{\cal P}_{\beta\gamma}+{\cal P}_{\beta\delta})
[\langle:\Phi_{\lambda_1}\Pi_\beta:\rangle_c
\langle:\Phi_{\lambda_2}\Phi_{\lambda_3}\Pi_\gamma\Pi_\delta:\rangle_c
+\langle:\Phi_{\lambda_1}\Phi_{\lambda_2}\Pi_\beta:\rangle_c
 \langle:\Phi_{\lambda_3}\Pi_\gamma\Pi_{\delta}:\rangle_c ]
\rbrace
\nonumber\\ \nonumber\\ &&
-6\lambda\Phi_0
(1+{\cal P}_{\alpha\beta}+{\cal P}_{\alpha\gamma}+
{\cal P}_{\alpha\delta})
\sum_{\lambda_1\lambda_2}
\langle\alpha|\lambda_1\lambda_2\rangle
(1+{\cal P}_{\beta\gamma}+{\cal P}_{\beta\delta})
\langle:\Phi_{\lambda_1}\Pi_\beta:\rangle_c
\langle:\Phi_{\lambda_2}\Pi_\gamma\Pi_\delta:\rangle_c
\nonumber\\ &&
+\frac{3}{2} \lambda
(1+{\cal P}_{\alpha\beta}+{\cal P}_{\alpha\gamma}+
{\cal P}_{\alpha\delta})
\sum_{\lambda} \langle\beta\gamma\delta|\lambda\rangle
\langle:\Phi_\lambda \Pi_\alpha:\rangle_c \; .
\label{lasteqom}
\end{eqnarray}
\section{The Gaussian effective potential for $\Phi^4_{1+1}$-theory}
\label{gep}
The Hamiltonian for the $\Phi^4_{1+1}$-system is given by
\begin{eqnarray}
H=\frac{1}{2} \int dx \left[ \pi^2 + \left(\frac{d\phi}{dx}\right)^2
+ m_0^2 \phi^2 + \frac{\lambda}{4} \phi^4 \right]
\end{eqnarray}
with $m_0^2=m^2+\delta m^2$, $\delta m^2=-3\lambda \Delta(0)$
(see section \ref{correldyn}).
Let furthermore $\phi=\Phi_0 + \Phi$,
$$ \Phi(x)=(2\pi)^{-\frac{1}{2}} \int dp \;
(2\omega_p)^{-\frac{1}{2}}
\left[ a_p + a_{-p}^\dagger \right] e^{ipx} \; , $$
$$
\omega_p=\sqrt{p^2+m^2}\; , \quad a_p |0 \rangle \;=0 \; ,
$$
where $|0 \rangle $ is the perturbative vacuum.
As in section \ref{correldyn}, $\Phi_0$ denotes the constant vacuum
magnetization, i.e. the ground state expectation value of $\phi$.

The GEP approximation consists in the ansatz
\begin{eqnarray}
H_0=\frac{1}{2} \int dx \left[ \pi^2 + \left(\frac{d\phi}{dx}\right)^2
+ M^2 \left( \phi - \Phi_0 \right)^2 \right]
\end{eqnarray}
for the variational Hamiltonian, i.e. the interacting system is
approximated
by a free system with an effective mass $M$, which serves as a
variational
parameter. One then has to minimize the expectation value of $H$ with
respect to the ground state of $H_0$.

This method is equivalent to a BCS calculation, where the variational
wavefunction is given by a boson pair condensate
\begin{eqnarray}
|\Psi \rangle \; ={\cal N} exp(-S) |0 \rangle \; , \quad
S=\frac{1}{2} \int dp \; \mu(p) a^\dagger_p a^\dagger_{-p} \; ;
\end{eqnarray}
here ${\cal N}$ denotes a normalization constant and $\mu(p)$ is the
variational parameter. $|\Psi \rangle $ is a quasiparticle
vacuum, i.e.
$b_p |\Psi \rangle =0$, where the $b_p$ can be obtained from the
$a_p$ by means of
a Bogoliubov-transformation.
We have
\begin{eqnarray}
\Phi(x)=(2\pi)^{-\frac{1}{2}} \int dp \; (2\Omega_p)^{-\frac{1}{2}}
\left[ b_p + b_{-p}^\dagger \right] e^{ipx}
\label{bmodenentwicklung}
\end{eqnarray}
with $\Omega_p=\omega_p \frac{1+\mu(p)}{1-\mu(p)}$.

After minimizing the energy functional, the GEP ansatz yields the
equation
\begin{eqnarray}
M^2-m^2=3\lambda \left\{ \Phi_0^2 + \int_{-\infty}^{\infty}
\frac{dp}{2\pi} \left( \frac{1}{2\Omega_p}
-\frac{1}{2\omega_p} \right) \right\}
\label{gapgleichung}
\end{eqnarray}
with $\Omega_p=\sqrt{p^2+M^2}$ for the effective mass $M$.
Equation (\ref{gapgleichung}) is identical to the
HFDS\footnote{{\bf H}artree {\bf F}ock {\bf D}yson {\bf S}chwinger}
equation and the Gap equation following from the BCS ansatz \cite{11}.

For known effective mass $M$, the equal-time Green functions can be
evaluated via (\ref{bmodenentwicklung}) as expectation values with
respect
to the quasiparticle vacuum $|\Psi \rangle $.
The value for the energy density functional as a function of $\Phi_0$
(obtained with the method described above) is the Gaussian
approximation
for the effective potential of the theory.
\end{appendix}
\newpage
\newpage
{\large \bf Figure Captions}
\newcounter{figno}
\begin{list}%
{\underline{fig.\arabic{figno}}:}%
{\usecounter{figno}\setlength{\rightmargin}{\leftmargin}}
\item
\label{picphstran0}
Vacuum magnetization as a function of the coupling in the
GEP approximation
for different numbers of basis states.
\item
\label{picconv}
Ground state energy density during the time-dependent process
as a function of the coupling for a given vacuum magnetization of
$\Phi_0=0.4$ and different values of $\beta$;
upper left: $\Phi^4CD(2)$, upper right: $\Phi^4CD(2,3)$,
lower left: $\Phi^4CD(2,4)$, lower right: $\Phi^4CD(2,3,4)$;
in addition the static GEP curve is displayed in the upper left plot.
\item
\label{picinccpl}
Ground state energy density as a function of the coupling for
different
vacuum magnetizations; all curves except the one in GEP
approximation have
been evaluated time-dependently with
$\beta=5 \; {\rm MeV}^2c/{\rm fm}$;
upper left: GEP, upper right: $\Phi^4CD(2,3)$,
lower left: $\Phi^4CD(2,4)$, lower right: $\Phi^4CD(2,3,4)$.
\item
\label{picveff}
Effective potential as a function of the vacuum magnetization
for different values of the coupling; all curves except the one in GEP
approximation have been evaluated time-dependently with
$\beta=5 \; {\rm MeV}^2c/{\rm fm}$;
upper left: GEP, upper right: $\Phi^4CD(2,3)$,
lower left: $\Phi^4CD(2,4)$, lower right: $\Phi^4CD(2,3,4)$.
\end{list}
\end{document}